\documentclass[journal]{IEEEtran}

\usepackage{xcolor,soul,framed} 
\usepackage{multirow}
\usepackage{graphicx}
\usepackage{amsmath}
\usepackage{ragged2e}
\usepackage{caption} 
\usepackage{booktabs}
\usepackage{hyperref}
\usepackage{booktabs}
\usepackage{mwe}
\usepackage{booktabs, makecell, multirow, tabularx}
\setcellgapes{2pt}
\newcolumntype{L}[1]{>{\RaggedRight\hspace{0pt}%
                     \hsize=#1\hsize}X}
\usepackage{lipsum}
\usepackage{atbegshi}
\AtBeginDocument{\AtBeginShipoutNext{\AtBeginShipoutDiscard}}

\begin{document}
\bstctlcite{IEEEexample:BSTcontrol}
    \title{Multichannel joint-polarization-frequency-modulation encrypted metasurface in secure THz communication}
  \author{Abbas Ozgoli, Parsa Farzin, Mohammad Javad Hajiahmadi*, and Mohammad Soleimani}

  \thanks{School of Electrical Engineering, Iran University of Science and Technology, Tehran, 1684613114, Iran (email:hajiahmadi@iust.ac.ir)}

\maketitle

\begin{abstract}
Since the discovery of wireless telegraphy, wireless communication via electromagnetic (EM) signals has become a standard solution to meet the growing demand for information transfer in modern society. To prevent counterfeiting and manipulation by unauthorized individuals and agencies, it is crucial to innovate and enhance security through information encryption. In this paper, we introduce a metasurface that controls amplitude modulation at two different frequencies. Here, focusing on amplitude-frequency modulation for both x- and y- polarizations, we present an encrypted wireless communication protocol that using the chaos algorithm to secure the target data and prevent eavesdroppers from accessing it. The encrypted data is transmitted through the varying amplitudes at two different frequencies for both linear polarizations, achieved using two distinct graphene layers individually controlled by external biasing conditions. The extensive freedom enabled by simultaneous modulation of amplitude, frequency, and polarization allows information to be transmitted across diverse channels, thereby enhancing the security of encrypted information. The simulations demonstrate that encoding data, such as images, and transmitting it through amplitude-frequency modulation for both horizontal and vertical polarizations offer promising opportunities for various applications, including THz communications, anti-counterfeiting, THz data storage, and THz data transmission.
\end{abstract}


%
\IEEEpeerreviewmaketitle


\section{Introduction}

\IEEEPARstart{T}{erahertz} (THz) waves, spanning 0.1-10 THz (residing between microwave and infrared wavelengths), present unmatched prospects for numerous crucial applications, including imaging\cite{wang2022overview}, sensing\cite{3}, and wireless communication\cite{4}. THz signals can harness the benefits of both millimeter-wave and optics, offering high spatial resolution and deep penetration into dielectric materials or human tissue without causing harmful ionization \cite{sun2011promising}. However, conventional THz devices are often constrained by complex configurations and bulky sizes, making them unsuitable for system integration and thereby limiting their further development. Developing compact, conformal, and simple devices is crucial to effectively control the propagation properties of free-space THz waves \cite{wang2022simple}. However, the remarkable properties of metamaterials have eliminated the challenges \cite{qureshi2024polarization}.

Throughout history, protecting information has been a vital component of successful communication. In today's digital era, it's essential to safeguard private data to combat concerns regarding data sharing and abuse. We can use different cryptographic techniques to secure stored information and prevent potential threats and tampering\cite{14,15}. As information security becomes increasingly important, numerous cryptographic techniques have emerged that manipulate the features of EM waves (such as amplitude, phase, polarization, frequency, and orbital angular momentum (OAM)) to encode and decode information \cite{zheng2021metasurface,li2020advanced}. This opens new possibilities for secure communication and data storage. Emerging from the field of computer science in 1882, cryptography revolutionizes the process of converting plaintext into encrypted ciphertext \cite{lin1996cryptography}. Given the critical importance of information security, numerous encryption and decryption techniques have been developed \cite{jacobs1996quantum}. Modern cryptography offers a robust array of methods to counteract malicious adversaries while ensuring legitimate users have access to the necessary information \cite{xomalis2019cryptography}. The potential to harness the vast information capacity of metasurfaces strongly motivates the development of advanced image encoding and information encryption techniques \cite{rajabalipanah2020real}. These innovations are particularly crucial for top-secret communications, where the secure exchange of confidential information is paramount.

\begin{figure*}[hbt!]
    \centering
    \includegraphics[scale=0.5]{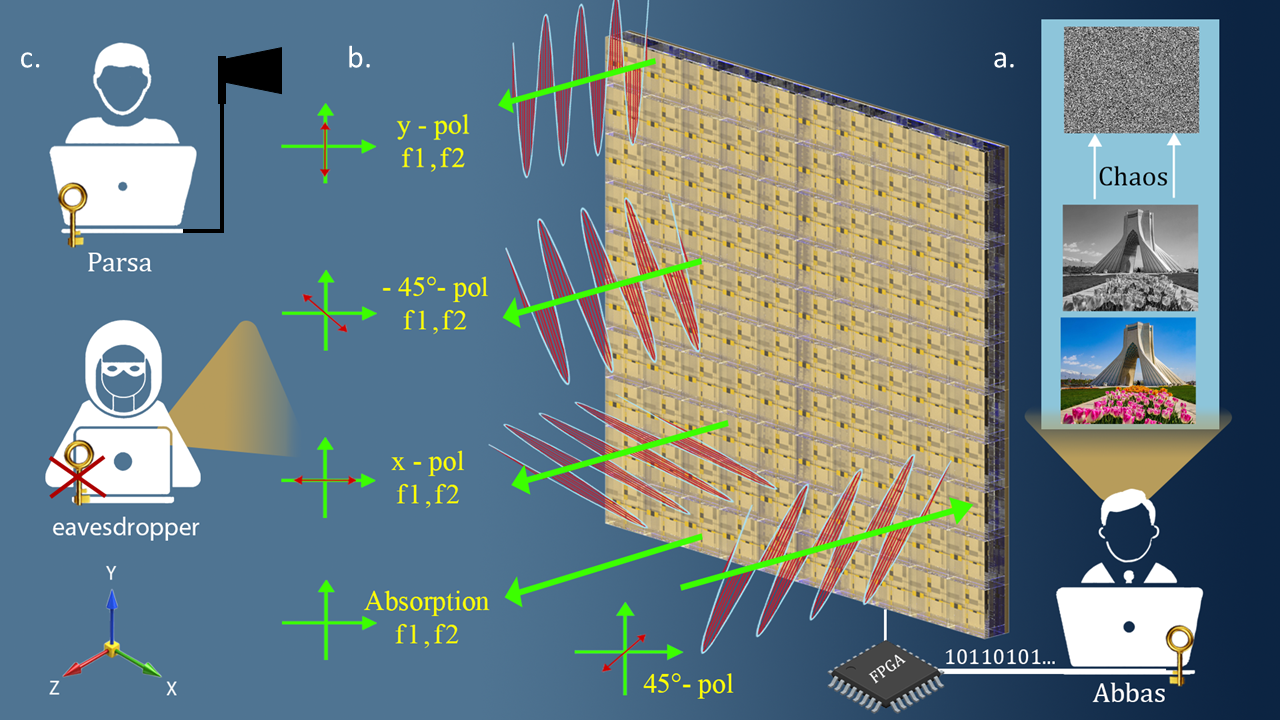}
    \caption{A conceptual illustration of the amplitude-frequency modulation metasurface for both horizontal and vertical polarizations. (a) Encrypting the target image, Azadi Tower, using the chaos algorithm and transmitting the generated bits to the FPGA. (b) Transmitting bits via FPGA to the metasurface for real-time manipulation of amplitude separately for horizontal and vertical polarizations at two different frequencies. (c) Receiving the information on the receiver side and decrypting it using the key.}
    \label{fig1}
\end{figure*}
 
The two-dimensional variant of metamaterial, known as a metasurface, consists of two-dimensional arrays of artificial atoms \cite{minovich2015functional,ghorbani2021deep}. These metasurfaces have garnered significant interest due to their compact nature and exceptional ability to manipulate electromagnetic waves (EM) at subwavelength scales \cite{yu2014flat,luo2019subwavelength}. Due to the precise design of each meta-atom, metasurfaces can independently or simultaneously modulate the amplitude, phase, and polarization of incident EM waves \cite{xu2021topology}. Consequently, numerous metasurfaces have been developed to enable extraordinary applications, such as holography \cite{yin2024multi}, meta-lenses \cite{rouhi2021multi, focusing_TR}, beam steering \cite{jahangiri2024beam,rouhi2018real}, orbital angular momentum generation \cite{qin2018transmission}, and cloaking \cite{chen2013nanostructured}. In a groundbreaking advancement, Cui et al. have introduced the revolutionary concept of digital metasurfaces, transforming the field by bridging the physical and digital worlds, where electromagnetic responses are defined by coding patterns that dictate the arrangement of coding particles within a 2D array \cite{cui2014coding}. This innovative approach provides a novel perspective on metasurfaces through the lens of information science. In the digital coding metasurfaces, parameters such as amplitude, phase, polarization, and frequency can be coded as digital states "0" or "1". Compared to traditional wave manipulation tools like transformation optics \cite{barati2018experimental}, digital metasurfaces offer the advantages of easier analysis, design, and fabrication. Additionally, they enable a broader range of wave-matter functionalities \cite{liu2017concepts}. To date, most metasurfaces designed for specific applications rely on static and passive designs, defined by geometrical parameters such as particle size, shape, and array layout \cite{liu2016anisotropic,liang2016broadband}. In fact, their operational statuses remain fixed. As a result, a major bottleneck currently hindering the advancement of this field is the lack of efficient devices that enable dynamic tunability of functionality over time \cite{rouhi2019multi}. Unlike traditional metasurfaces, reprogrammable metasurfaces with well-defined elements can be programmed and digitized using a field-programmable gate array (FPGA) \cite{farzin2024reprogrammable}. 

On the other hand, polarization manipulation of EM waves to generate multiple polarization-dependent channels and accomplish diverse functionalities without incorporating additional devices is a crucial issue in EM wave research \cite{farzin2023graphene,he2021multifunctional,sun2019electromagnetic}. Polarization modulation, which uses the polarization mode of EM waves as an information-bearing parameter, has been proposed as a suitable alternative to conventional modulation techniques such as amplitude-shift keying, frequency-shift keying, and phase-shift keying \cite{huang2021polarization}. Therefore, in wireless communication, the use of these metasurfaces enables the transmission of encrypted information through different polarizations \cite{wang2022polarization,ma2020editing,wang2023polarization}. Although real-time amplitude modulation at several frequencies has been achieved in the microwave band \cite{hong2021programmable,wu2023multi}, it faces numerous limitations for wireless communication. These limitations include a very low amplitude-frequency modulation bit count and a low data transmission rate due to the use of the microwave band.

\begin{figure*}
    \centering
    \includegraphics[scale=0.5]{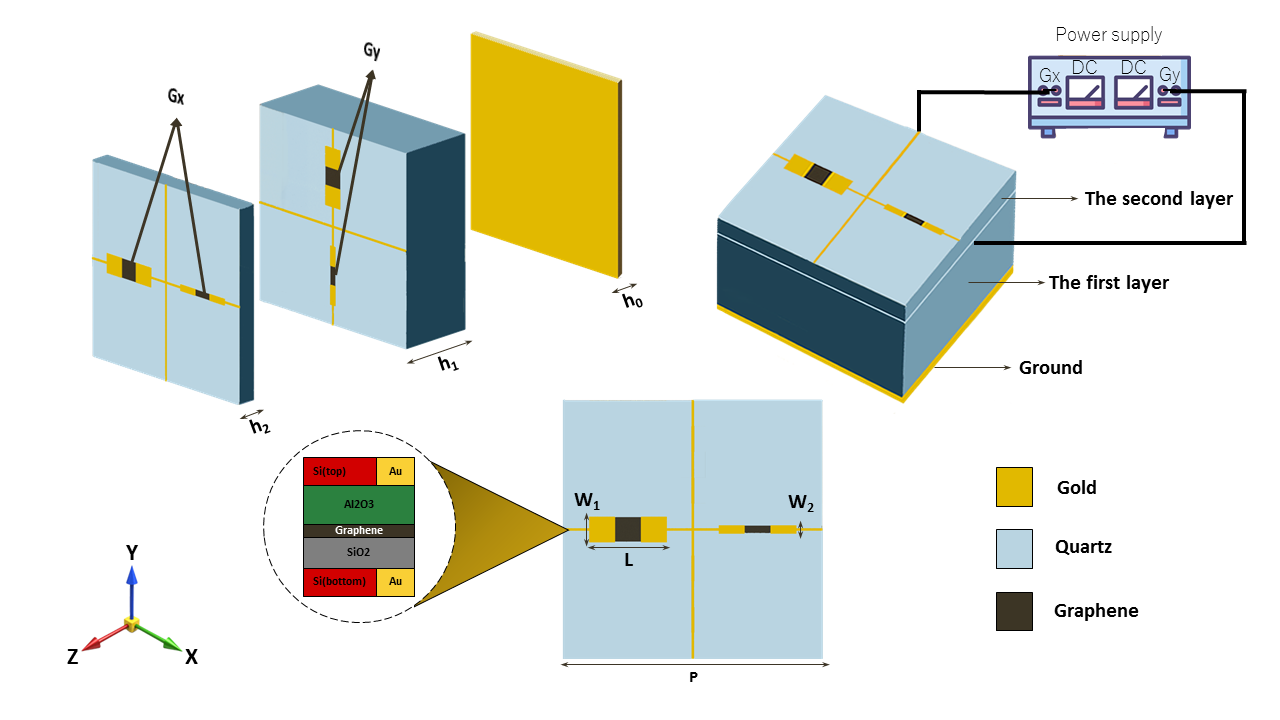}
    \caption{The proposed reprogrammable meta-atom dimensions with details and from different sights.
}
    \label{fig2}
\end{figure*}

In our previous research \cite{farzin2024multi}, We successfully designed a metasurface in the THz frequency band using graphene, capable of real-time modulation of incident wave polarization to achieve a 2-bit metasurface, enabling the transmission of encrypted data via DRPE through different polarization channels at a single frequency. One important limitation of this work is the low bit count, as it operates on just one frequency channel for data encryption, which decreases the security of the communication channel. In this paper, We introduce a programmable metasurface capable of controlling the reflected THz wave amplitude at two different frequencies for both x- and y-polarizations separately and in real time. Furthermore, proposed applications for securing communication use various encryption schemes, where a chaos algorithm protects the transmitted information. The proposed meta-atom is composed of a three-layered structure consisting of a gold layer and two quartz layers. Each quartz layer contains a combination of large and small rectangles made of graphene and gold. By precisely adjusting the fermi level energy within each graphene double-layer, the propose metasurface achieves real-time control over the amplitude of both linear polarizations at two distinct frequencies. The metasurface transmits encrypted information across different polarization amplitude in different frequency channels by adjusting the Fermi energy level of the graphene layers using FPGA. The metasurface utilizes four distinct amplitude levels for each x- and y-polarization at two different frequencies, resulting in a 4-bit meta-atom. The ability to simultaneously modulate amplitude, frequency, and polarization provides extensive flexibility, allowing encrypted information to be transmitted through multiple channels. This, in turn, enhances security by increasing channel capacity. The results shows that the proposed writable information metasurface achieves exceptional performance in amplitude-frequency modulation for both x- and y-polarizations, significantly enhancing information security. This innovative metasurface offers a high degree of flexibility for applications like  THz data storage, THz data transmission, and THz communications.

\section{Reconfigurable Amplitude Modulation using Information Metaurface}

Encryption using secret keys, a fundamental aspect of cryptography, enables secure communication and ensures that an encrypted message sent over a public channel remains incomprehensible to any unauthorized third party\cite{16,17}. Figure 1 illustrates the schematic of secure multichannel wireless communication using a frequency-amplitude modulation metasurface for different polarizations in the THz band. As depicted in Figure 1a, Abbas (the sender) has encrypted the desired color image, the Azadi Tower in Tehran, using the chaos algorithm. The encrypted information is converted into binary bits, which are then transferred to the FPGA. In Figure 1b, the encrypted information is transferred to the metasurface, resulting in amplitude modulation across two different frequency bands for both horizontal and vertical polarizations. The encrypted information is then transmitted through various channels of polarization. Encrypting the information ensures that even if an eavesdropper gains access to it, they will be unable to understand the original data since they do not possess the key. It is important to note that the key is exchanged between the sender and receiver through a private channel. On the receiver side, as shown in Figure 1c, Parsa (the receiver) receives the transmitted information using an antenna sensitive to both frequencies and both horizontal and vertical polarizations. Since the key is only held by the sender and the receiver, Parsa can decode the received information and ultimately view the transmitted color photo of Azadi Tower.

\begin{figure*}
    \centering
    \includegraphics[scale=0.5]{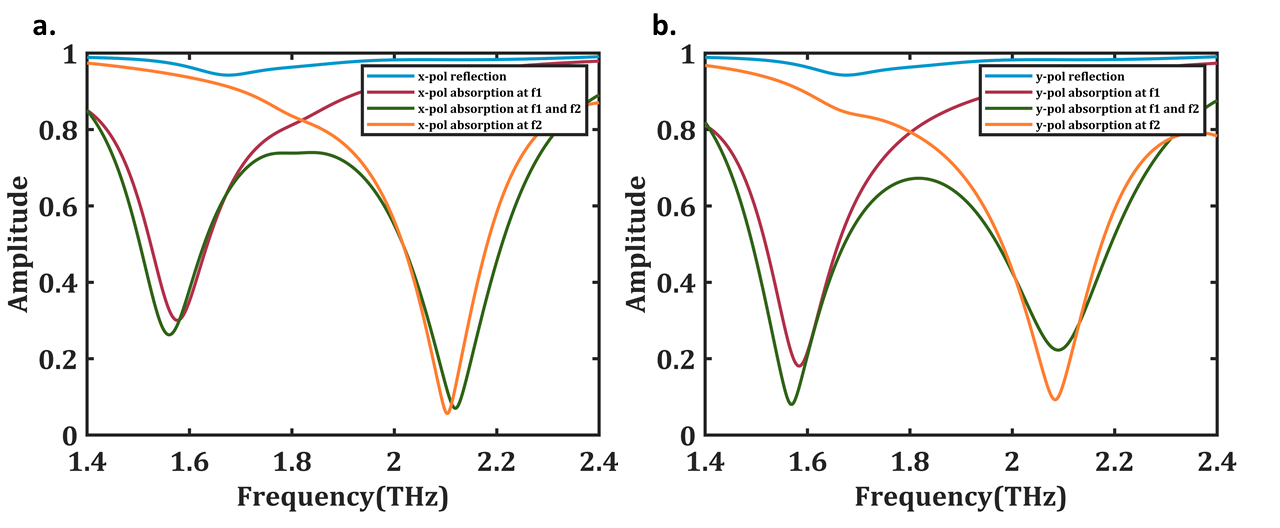}
    \caption{Amplitude responses of the x- and y-polarizations, exhibiting two resonances of the same frequencies on both x-polarization (a) and y-polarization (b).
}
    \label{fig3}
\end{figure*}

\section{Materials and Methods}
\subsection{Graphene's surface conductivity} 

To define Graphene properties, one should use a complex surface conductivity tensor that reflects insights from quantum mechanical principles, micro-level physical analysis, and semi-classical theory. This conceptualization is an advancement of the dual-sided conductivity model initially developed for carbon nanotubes and has been specifically adapted to address the conductive behavior of isotropic graphene\cite{19}.
\begin{equation}
   \sigma(\omega, \mu_c(E_0 ), \Gamma, T, B_0 )= \hat{x}\hat{x}\sigma_{xx} + \hat{x}\hat{y}\sigma_{xy} + \hat{y}\hat{x}\sigma_{yx} + \hat{y}\hat{y}\sigma_{yy}
   \label{eq1}
\end{equation}

Here, $\omega$ signifies the angular frequency, $\mu_c$ is the chemical potential, $\Gamma$ represents a constant electron scattering rate unrelated to energy (called a phenomenological scattering rate), $T$ refers to the absolute temperature, and $B_0 = \mathop z\limits^ \wedge  B_0$ indicates a magnetostatic bias field that applies in the direction of the z-axis\cite{20}. The surface conductivity of a graphene monolayer, represented by $\sigma(\omega)$, captures its optical attributes. This conductivity is computed from the Kubo formula, combining both the intra-band component $\sigma^{\text{intra}}(\omega)$ and the inter-band component $\sigma^{\text{inter}}(\omega)$\cite{21,tahmasebi2022parallel}, is given by

\begin{equation}
\sigma_{\mathrm{g}} = \sigma^{\mathrm{intra}}_{\mathrm{g}}+\sigma^{\mathrm{inter}}_{\mathrm{g}},
\label{eq: GraphCond}
\end{equation}

\begin{equation}
    \sigma^{\text{intra}}_{\mathrm{s}} (\omega, \mu_c, \Gamma, T)=-j\frac{qe^2 k_{\text{B}} T}{\pi \hbar^2 (\omega-j\ 2\Gamma)}
    [\frac{\mu_c}{k_{\text{B}} T} + 2ln(e^\frac{-\mu_c}{k_B T}+1)]
    \label{eq3}
\end{equation}

The inter-band surface conductivity can be approximated for ${k_{\text{B}}}T \ll \left| {{\mu _c}} \right|,\omega h$ as:

\begin{equation}
    \sigma^{\text{inter}}_{\mathrm{s}} (\omega, \mu_c, \Gamma, T) \approx -j \frac{qe^2}{4\pi \hbar} \left(\frac{2|\mu_c |-(\omega-j/\tau)\hbar}{2|\mu_c |+(\omega-j/\tau)\hbar}\right)
   \label{eq4}
\end{equation}

Here, ${q_e}$ is electron charge, $h$ reduced Planck constant, ${K_{\text{B}}}$ is Boltzmann’s constant, $\tau$ is relaxation time, and $\Gamma  = 1/(2\tau )$ is the scattering rate, represents how it loses energy. At frequencies below 8 THz, the conductivity hinges on ${\mu _c}$, which can be altered by modifying carrier density through gating or doping, thus affecting graphene’s conductive properties. Using the Drude model, this behavior is simplified to focus on the density of charge carriers and their mobility, ensuring accurate application when electron scattering is frequent within the small spatial confines of graphene samples \cite{22,23}. In this study, we consider $T$ = 300~K and $\tau$ = 1 ps and Graphene thickness ($t_{\text{g}}$) = 5~nm throughout our research. It’s important to highlight that the surface conductivity of the graphene surface can be dynamically modified by an external electrical bias (Further elaboration available in Supplementary A), thus enabling real-time implementation of various functions for the proposed metasurface as depicted in Figure 1. 

\begin{table*}
\caption{\label{1} Possible 4-bit situations according to amplitudes, polarizations, and frequencies.}
\centering
\begin{tabular}{@{}lllll@{}} 
\toprule
\textbf{No.} & \textbf{State} & \textbf{Gx} & \textbf{Gy} & \textbf{Feature}                                     \\ \midrule
\textbf{0}   & 0000            & 0.58           & 0.8           & Absorption at both frequencies and   polarizations   \\ \midrule
\textbf{1}   & 0001            & 0.31        & 0           & x-pol absorbs at $f_1$ and y-pol reflects               \\ \midrule
\textbf{2}   & 0010            & 0.58        & 0           & x-pol absorbs at $f_1$ and $f_2$ and y-pol   reflects      \\ \midrule
\textbf{3}   & 0011            & 1.09        & 0           & x-pol absorbs at $f_2$ and y-pol reflects               \\ \midrule
\textbf{4}   & 0100            & 0           & 0.44        & y-pol absorbs at $f_1$ and x-pol reflects               \\ \midrule
\textbf{5}   & 0101            & 0           & 0.8         & y -pol absorbs at $f_1$ and $f_2$ and x-pol   reflects     \\ \midrule
\textbf{6}   & 0110            & 0           & 1.44        & y -pol absorbs at $f_2$ and x-pol   reflects            \\ \midrule
\textbf{7}   & 0111            & 0.31        & 1.44        & x-pol absorbs at $f_1$ and y-pol absorbs   at $f_2$        \\ \midrule
\textbf{8}   & 1000           & 1.09        & 0.44        & y-pol absorbs at $f_1$ and x-pol absorbs   at $f_2$        \\ \midrule
\textbf{9}   & 1001           & 0.31        & 0.8         & x-pol absorbs at $f_1$ and y-pol absorbs at $f_1$ and $f_2$   \\ \midrule
\textbf{10}  & 1010           & 0.58        & 0.44        & y-pol absorbs at $f_1$ and x-pol absorbs   at $f_1$ and $f_2$ \\ \midrule
\textbf{11}  & 1011           & 1.09        & 0.8         & x-pol absorbs at $f_2$ and y-pol absorbs   at $f_1$ and $f_2$ \\ \midrule
\textbf{12}  & 1100           & 0.58        & 1.44        & y-pol absorbs at $f_2$ and x-pol absorbs   at $f_1$ and $f_2$ \\ \midrule
\textbf{13}  & 1101           & 1.09        & 1.44        & -45°-pol at $f_1$ reflects                              \\ \midrule
\textbf{14}  & 1110           & 0.31        & 0.44        & -45°-pol at $f_2$ reflects                              \\ \midrule
\textbf{15}  & 1111           & 0        & 0         & Reflection at both frequencies and   polarizations   \\ \bottomrule

\end{tabular}
\end{table*}

\subsection{Unit-cell Design} 

The proposed unit-cell is a structure comprising multiple resonating elements to achieve broader bandwidths and the structure employs gold, graphene and a quartz dielectric with a relative permittivity (${\varepsilon _r}$) of 3.75 and a loss tangent ($tand$) of 0.0004. This metasurface includes graphene and gold combined in two different big and small rectangle shapes over two quartz substrates with thicknesses of $h_{1}=20$µm  and $h_{2}=0.3$µm and a gold ground plane with thickness of $h_{0}=0.2$µm to have a perfect reflection. By positioning the rectangles vertically in the first layer and horizontally in the second layer, we can determine the amplitudes of the y- and x-polarization states, respectively. The purpose of using small and large rectangles of graphene and gold in each layer is to regulate the amplitude at various frequencies. This alteration in the dimensions of graphene and gold modifies their equivalent circuit (more information in supplementary C), resulting in resonance occurring at different frequencies when a chemical potential is applied to graphene. The periodicity of the meta-atom is $P = 50$ µm. The width of the large and small rectangle is $W_{1}$= 6~µm and $W_{2}$=2~µm, respectively and both rectangles have a length of $L = 16$~µm. Figure 2 illustrates the structure of the unit cell, detailing each layer along with the corresponding parameter values and the biasing method. The proposed biasing technique for the metasurface offers notable convenience, as the biasing of a single meta-atom allows for the simultaneous biasing of all meta-atoms within the metasurface. The metasurface design was achieved through numerical simulations using the commercial EM software CST Microwave Studio. We applied $x-y$ boundary conditions for unit cell design and open space boundary conditions for $+z$ and $-z$ directions with Floquet port. As a result, it is possible to independently and dynamically manipulate the amplitude to represent "0" and "1" bits for each polarization, enabling the attainment of a 4-bit polarization state.

Previous research predominantly utilized a simple capacitive structure involving graphene, an insulator, and a metallic electrode to modulate graphene's conductivity \cite{kim2018amplitude}. However, this method requires a continuous external voltage, leading to inevitable static power consumption. To overcome this limitation, we propose a graphene-based multilayer structure with minimal static power consumption, leveraging a non-volatile floating-gate graphene configuration commonly used in non-volatile devices. This structure comprises Si, $SiO_{2}$, $Al_{2}O_{3}$, and single-layer graphene. Our floating-gate design adjusts the charge density of graphene. When we apply a positive voltage to the top Si layer, electrons tunnel through the $SiO_{2}$ layer and are captured by the graphene, increasing its charge density\cite{momeni2022switchable}. Conversely, applying a reverse voltage allows electrons from the graphene to tunnel through the $SiO_{2}$ layer back to the lower Si layer, reducing the charge density. Due to the electrical isolation of the graphene from the Si layers, the charge density remains stable even after the voltage is disconnected, thus requiring no additional power. The $SiO_{2}$ and $Al_{2}O_{3}$ layers are 10 nm and 20 nm thick, respectively. Electron tunneling occurs exclusively through the thinner $SiO_{2}$ layer, as the thicker $Al_{2}O_{3}$ layer prevents tunneling. The relative permittivity of $Al_{2}O_{3}$ is 9, and that of $SiO_{2}$ is 3.9. These ultra-thin layers are critical for DC bias design, though their effect on the amplitude and phase of reflected EM waves is negligible in THz simulations\cite{rajabalipanah2020real}.

The procedure for fabricating the sample can be detailed, demonstrating a viable approach using existing fabrication techniques. First, a layer of quartz covers a silicon wafer via a spin-coating solution. Next, to encourage the development of an ultrathin, high-quality $SiO_{2}$ tunnel oxide, p-Si wafers are placed in a Rapid Thermal Oxidation (AS-One) chamber at 25°C with a nitrogen flow of about 800 sccm. The temperature is then elevated to 900°C at approximately 25°C/s, maintained to grow the $SiO_{2}$ with an oxygen flow of around 800 sccm at 900°C for 90 seconds. Afterward, the temperature is gradually lowered from 900°C to 25°C at a rate of about 3°C/s under a nitrogen flow of approximately 800 sccm\cite{farzin2024multi}. Graphene sample preparation uses chemical vapor deposition (CVD) on copper foil. Once transferred to $SiO_{2}$/poly-Si wafers, the surplus graphene is eliminated using 100 keV electron beam lithography in PMMA, followed by an oxygen plasma etch, leaving only the desired square or ribbon-shaped regions. Subsequently, a 20 nm thick layer of $Al_{2}O_{3}$ is deposited via atomic layer deposition (ALD), utilizing trimethyl aluminum (TMA) and $H_{2}O$ as precursors at 200°C. Lastly, another layer of quartz is applied to the silicon wafer using a spin-coating solution and placed on top of the $Al_{2}O_{3}$ layer. This process gets repeated two times, with each layer added sequentially\cite{rouhi2019multi}.

\begin{table*}
\caption{\label{2} The method of moving each pixel of image}
\centering
\begin{tabular}{lllllllllll}
\hline
Row number                & 1    & 2    & 3    & 4    & 5    & 6    & 7    & 8    & 9    & 10   \\ \hline
Chaotic generated numbers & 0.64 & 0.96 & 0.06 & 0.87 & 0.18 & 0.38 & 0.60 & 0.44 & 0.04 & 0.91 \\ \hline
Sorting numbers           & 0.04 & 0.06 & 0.18 & 0.38 & 0.44 & 0.60 & 0.64 & 0.87 & 0.91 & 0.96 \\ \hline
Row displacement          & 9    & 3    & 5    & 6    & 8    & 7    & 1    & 4    & 10   & 2    \\ \hline
\end{tabular}
\end{table*}

\section{Results}

Graphene can exhibit different electrical properties based on its Fermi energy (${\mu _c}$). When the Fermi energy is low, graphene acts as a dielectric, meaning it does not conduct electricity well due to a limited availability of charge carriers \cite{farzin2023graphene}. However, as the Fermi energy increases, graphene transitions to a conductive state, where the electrons have enough energy to move freely, making graphene an excellent conductor \cite{farzin2023graphene}. This transition from a dielectric to a conductor occurs because the higher Fermi energy increases the number of charge carriers, allowing for efficient electrical conduction. We denote the chemical potential assigned to the graphene in the first layer, which is associated with y-polarization, as $\mathrm{G_y}$ and the chemical potential assigned to the graphene in the second layer, which is associated with x-polarization, as $\mathrm{G_x}$. Here ${\mu _c}=\{ \mu _c^{\mathrm{G_x}},\mu _c^{\mathrm{G_y}}\}$, with $\mu _c^{\mathrm{G_x}}$ and $\mu _c^{\mathrm{G_y}}$ indicating the chemical potential applied to $\mathrm{G_x}$ and $\mathrm{G_y}$, respectively. As shown in Figure 1, the metasurface is subjected to wave radiation with a 45°-polarization, enabling control over the polarization amplitude in both the x- and y-polarization directions simultaneously. It is important to note that at this 45°-polarization, the amplitude and phase are identical. As a result, by only varying the mentioned chemical potential in different values, as outlined in Table 1, and keeping all other parameters constant, the structure can dynamically and simultaneously adjust the strengths of x- and y-polarization of a slant incident wave at two distinct frequencies, $f_1$ (1.57 THz) and $f_2$ (2.1 THz) in real-time. The chemical potential values for each $\mathrm{G_x}$ and $\mathrm{G_y}$ are as follows: \(\left[ \left\{ 0\,\text{eV},\,0.31\,\text{eV},\,0.58\,\text{eV},\,1.09\,\text{eV} \right\}, \left\{ 0\,\text{eV},\,0.44\,\text{eV},\,0.8\,\text{eV},\,1.44\,\text{eV} \right\} \right]\). Referring to Figure 3 and Table 1, we identify 16 distinct states that display eight amplitude responses, with four for each polarization. Each amplitude response of a given polarization can simultaneously occur with any of the four amplitude responses of the other polarization. The two categories of chemical potential considered for the metasurface exhibit increasing values. Initially, when the chemical potential value is zero, the metasurface reflects the desired polarization. Next, increasing the chemical potential to the first value induces resonance at frequency $f_1$. Further growing it to the second value causes resonance at both $f_1$ and $f_2$. At the highest value, resonance occurs exclusively at $f_2$. (Additional details of full structure results are provided in Supplementary B). According to the defined states, one could transmit information (specifically images) across various channels.

\section{Multi-channel Cryptography Using the Proposed Meatasurface}
There are three primary types of image encryption algorithms: those that change the position of the image, those that alter the values of the pixels, and those that use vision transformation\cite{24}. Our project involves implementing image encryption and decryption utilizing a 3D logistic chaos function.
The essence of chaos theory is discovering order within disorder. It means searching for orders not only on a small scale but also on a larger scale. Sometimes, a phenomenon may seem completely random and unpredictable on a local scale, but it could exhibit a certain degree of regularity and predictability on a larger scale. Due to their seemingly chaotic nature, high sensitivity to initial conditions, and fast computational speed compared to other algorithms, these functions are used in image encryption. Different chaos functions are as follows: the Lorenz, Rossler, Logistic, Henon, Tent, and Baker functions, as well as others\cite{25,26}.
This algorithm uses the sequences generated by the three-dimensional logistic chaos function and XOR operation. The function produces three sequences of random numbers between zero and one using initial conditions and parameters. These sequences are encrypted and then utilized in the algorithm.

The three-dimensional logistic function is defined as follows and produces a chaotic sequence for the specified parameters\cite{li2019image}:

\begin{equation}
   x_{i+1}= \lambda x_i(1-x_i) + \beta y_i^2 x_i + \alpha z_i^3,   \qquad \qquad 3.53<\lambda<3.81 
\end{equation}
\begin{equation}
    y_{i+1}= \lambda y_i(1-y_i) + \beta z_i^2 y_i + \alpha x_i^3,   \qquad \qquad 0<\beta<0.022 
\end{equation}
\begin{equation}
    z_{i+1}= \lambda z_i(1-z_i) + \beta x_i^2 z_i + \alpha y_i^2,   \qquad \qquad 0<\alpha<0.015
\end{equation}

Following the initial image reading, a preliminary assessment determines if resizing to a 256 $\times$ 256 dimension is warranted. Subsequently, to optimize processing efficiency, one of the three color channels is selectively utilized to convert the image to grayscale, laying the groundwork for subsequent stages\cite{zhang2023chaos}.

We select n numbers from the sequence x generated between zero and one and then multiply these numbers by a thousand to obtain n numbers between one and one thousand. If the resulting number is even, shift the image matrix one column right, and if it is odd, move it one row down. In the next step, we will modify the pixel values using the XOR operation. XOR is a binary operation that returns a value of one only when one of the inputs is one. But not both. Otherwise, the output is zero$(A \oplus B = A\overline B  + \overline A B)$. We keep the first and second columns of the image matrix constant. XOR the first and third columns and place the result in the third column. We XOR the fourth column of the result with the sixth column. We put the result in the sixth column, and so on. We continue this operation until the final column. To change the matrix elements, we followed a specific process. We left the first and second columns unchanged and focused on the rest. To achieve this, we chose a number from the y sequence and assigned it to both the first and second columns of the matrix. After that, we generated random numbers between one and 256 and multiplied them by 255. After performing an XOR operation on the numbers with the first and second columns, we insert the resulting values into their respective columns\cite{chen2004symmetric}. It first converts the matrix of size $(m \times n)$ into a vector of size $(m \times n) \times 1$. Next, it selects a sequence of numbers z with the same number of terms as there are elements in the matrix and forms another vector. It then sorts the elements of the second vector in ascending order. The positions of these sorted elements in the second vector correspond to the displacement locations of the elements in the first vector. For instance, if the 50th element in the second vector is now at the 75th position after sorting, it moves the 50th element in the first vector to the 75th position, similarly for all other elements. After moving the $(m \times n) \times 1$ vector, we convert the first one into the m*n matrix, which is the encrypted image. Table 2 shows how to change the place after sorting the generated numbers of the chaos function. For example, After sorting, the smallest number in row 9 (0.04) is moved to row 1. eventually, by limiting the gray color spectrum to 16 colors, we have adapted the encryption and the structure to each other\cite{hossain2014new} (Check supplementary C for colorful encryption).

\begin{figure*}
    \centering
    \includegraphics[scale=0.5]{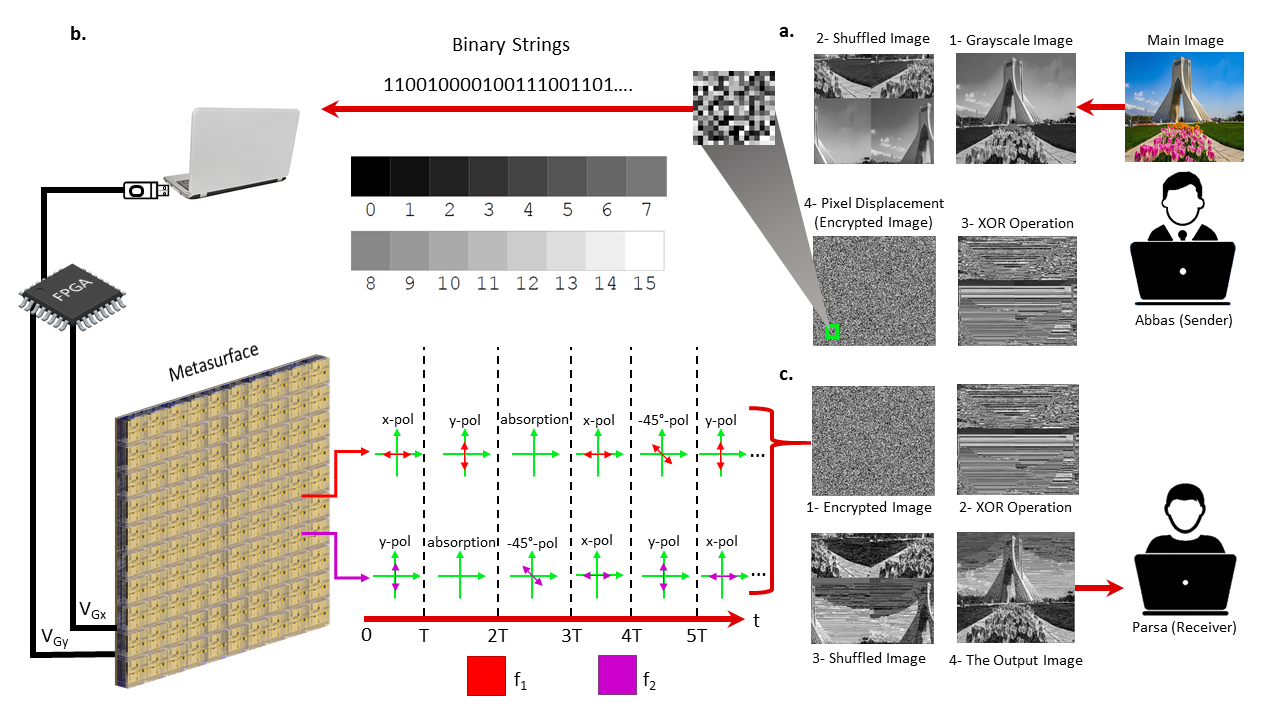}
    \caption{(a) Abbas (the sender) encrypts the desired photo, Azadi Tower, using the chaos algorithm, which produces 16 bits assigned to black to white colors and transfers the generated bits to the FPGA. (b) The bits are transferred to the metasurface via FPGA, and the encrypted information is transmitted through frequency channels and various polarizations at T-second intervals. (c) On the receiver side (Parsa), the information is received using special antennas and decrypted using a key. }
    \label{fig4}
\end{figure*}

Let us consider the following scenario. As shown in Figure 4, in order to establish secure communication without stealing information through eavesdropping, Abbas wants to send his target information, which is a color photo of Azadi Tower, to Parsa securely. First, he encrypts the target image using chaotic algorithm, generating a sequence of $256\times256=65536$ consisting of different colors. As illustrated in Figure 4, the spectrum of black and white colors is divided into 16 sections with different colors, in which case each color represents a special binary number (As shown in Table 1, state 0 is the black color represented by "0000" and state 15 is the white color represented by "1111"). The generated bits, assigned to their respective colors, are transferred to the FPGA via the computer. This transfer enables the FPGA to apply the necessary voltages to achieve the desired chemical potential in each horizontal and vertical graphene layer. These potentials are related to the amplitude of x- and y-polarizations at various frequencies, resulting in the desired bits being obtained. Consequently, based on the generated binary sequences, these bits are transmitted at various times t seconds through different polarizations with distinct amplitude levels across two frequency channels, $f_1$ and $f_2$. Expanding the number of channels and increasing the freedom of operations (achieved through amplitude modulation, polarization modulation, and frequency modulation) ensures robust encryption of information. This makes it difficult for eavesdroppers to decrypted the information if they gain access to the encrypted data. On the receiver side (Parsa), thanks to the antenna sensitive to both horizontal and vertical polarization and operating across frequencies $f_1$ and $f_2$, it effectively receives the information sent from the transmitter side. The decryption key is securely transmitted from the sender to the receiver through a private channel. Using the corresponding key, Parsa can decode the received information using the same chaotic algorithm employed on the sender's side. In this scenario, if an eavesdropper gains access to the information, they will be unable to decrypt it without the key. On receiver's side (parsa), decryption the received information reveals the final message: Azadi Tower.

\section{Conclusion}

In this paper, we introduced a reprogrammable metasurface capable of dynamically controlling amplitude modulation for both x- and y-polarizations at two different frequencies. The proposed meta-atom comprises three layers: a double layer of quartz and a gold layer serving as a reflector for the incident EM wave. Each quartz layer contain small and large rectangles made of gold and graphene are arranged. By applying an external electric voltage to each layer, the reflected wave is dynamically modulated in real time in terms of both amplitude and frequency for both x- and y-polarizations. Using the proposed metasurface, we encode a color image of the Azadi Tower using the chaos algorithm, and transmit the encoded information using amplitude-frequency modulation through both horizontal and vertical polarizations. The extensive freedom of operation provided by the proposed metasurface enables information to be transmitted across diverse channels, thereby enhancing the security of encrypted information compared to previous research. The proposed metasurface shows great potential in diverse applications, including THz data transmission, THz communications, and anti-counterfeiting.

\ifCLASSOPTIONcaptionsoff
  \newpage
\fi

\bibliographystyle{IEEEtran}
\bibliography{IEEEabrv,Bibliography}

\vfill

\end{document}